\documentclass{epl}

\title{Dephasing and Measurement Efficiency via a Quantum Dot Detector }%
\author{Gyong Luck Khym  \and Youngnae Lee  \and Kicheon Kang }

 \institute{
  \inst{} Department of Physics and Institute for Condensed
Matter Theory, Chonnam National University, Gwangju 500-757,
Korea\\    }

 \shortauthor{Khym {\em et al.} }
 \shorttitle{Dephasing via QDD}
\begin{document}
\maketitle

\begin{abstract}
We study charge detection and controlled dephasing of a
mesoscopic system via a quantum dot detector (QDD), where the
mesoscopic system and the QDD are capacitively coupled. The QDD is
considered to have coherent resonant tunnelling via a single
level. It is found that the dephasing rate is proportional to the
square of the conductance of the QDD for the Breit-Wigner model,
showing that the dephasing is completely different from the shot
noise of the detector. The measurement rate, on the other hand,
shows a dip near the resonance. Our findings are peculiar
especially for a symmetric detector in the following aspect: The
dephasing rate is maximum at resonance of the QDD where the
detector conductance is insensitive to the charge state of the
mesoscopic system.  As a result, the efficiency of the detector
shows a dip and vanishes at resonance, in contrast to the
single-channel symmetric non-resonant detector that has always a
maximum efficiency.  We find that this difference originates from
a very general property of the scattering matrix: The abrupt phase
change exists in the scattering amplitudes in the presence of the
symmetry, which is insensitive to the detector current but {\em
stores} the information of the quantum state of the mesoscopic
system.
\end{abstract}

 Suppression of the quantum interference due to
detection of a particle's path is a fundamental issue for
understanding the complementarity principle in quantum
theory~\cite{feynman65,stern90,scully91}. Recently, mesoscopic
physics is evolving into a stage where understanding the
measurement process becomes important. Of particular interest,
controlled dephasing of resonant tunnelling through a quantum dot
(QD) has been performed experimentally~\cite{buks98, sprinzak00,
kalish04}. In the experiment, a quantum point contact (QPC)
circuit electrostatically coupled to the QD enables detection of
the charge state of the QD, and accordingly suppresses the
coherent transmission of electrons through the QD. One may use
another kind of sensitive charge detector composed of a single
electron transistor (SET)~\cite{mahklin01,lu03}.  There exists the
trade-off between the measurement induced dephasing and the
information acquisition by the
detector~\cite{korotkov01,pilgram02}. The noise properties and
efficiency of a detector composed of a resonant-level conductor
has been investigated in Refs.~\cite{averin00,clerk04}.

Here we consider a fully phase-coherent `quantum dot detector'
(QDD) coupled to another quantum dot regarded as a `system'
(labelled as `QD-$s$', see Fig.~1)  with two possible charge
states, namely `0' and `1'. The detector is also composed of a
quantum dot attached to two electrodes (labelled as `QD-$d$') in
the coherent resonant tunnelling limit. Two quantum dots (QD-$s$
and QD-$d$) are capacitively coupled and electron transfer between
the two dots are forbidden. For simplicity, we assume that the
transmission of electron in the QDD takes place via a single
resonant level, which can be realized in the GaAs-based
two-dimensional electron gas (2DEG).
%%%%%%%%%%%%%%%%%%%%%%%%%%%%%%%%%%%%%%%%%%%%%%%%%%%%%%%%%%%%%%%%%%%%%%%%%%%%%%%
\begin{figure}[b]
\begin{center}%[width=5.0cm]
\resizebox{6cm}{!}{\includegraphics [11,37][240,125]{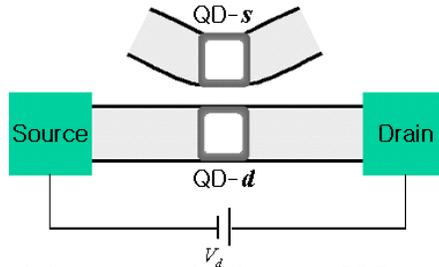} }
\caption{\label{Schematic} Schematic diagram of the quantum dot
detector (QDD) coupled to another quantum dot QD-$s$. The quantum
dot QD-$d$ of the QDD and QD-$s$  are capacitively coupled.}
\end{center}
\end{figure}
%%%%%%%%%%%%%%%%%%%%%%%%%%%%%%%%%%%%%%%%%%%%%%%%%%%%%%%%%%%%%%%%%%%%%%%%%%%%%%%
Our major observation in this study is the peculiar role of the
symmetry in the detector efficiency. We show that the detector
efficiency is reduced near the resonance even in a detector with
perfect time-reversal and mirror-reflection symmetry. This is in
contrast with the case of the symmetric single-channel
non-resonant detectors which have the maximum efficiency
independent of its transparency~\cite{korotkov01,pilgram02}. Based
on a general symmetry argument for the scattering matrix, we show
that the anomaly of a detector with resonance originates from the
abrupt phase change of the scattering amplitudes. This anomaly of
a resonant detector has not been taken into account
previously~\cite{averin00,clerk04}. We also discuss the relation
between the dephasing rate and the shot noise of the detector, and
the detector properties in the presence of the Fano resonance.

The Hamiltonian of the system under consideration is given by $ H
= H_d + H_{s} + H_{int}$, where $H_d$, $H_{s}$, and $H_{int}$
represent the QDD, the `system' containing QD-$s$, and the
interaction between the two subsystems, respectively. The
Hamiltonian for the QDD is expressed as
\begin{eqnarray}
 H_d =  \sum_{\alpha=R,L}\sum_{k } \varepsilon_{k}c_{\alpha k}^{\dag}c_{\alpha k}
 + \varepsilon_{d} d^{\dag}d
  +  \sum_{\alpha=R,L}\sum_{k} \big( V_{\alpha} d^{\dag}c_{\alpha k }  +
 h.c. \big)~, \label{QDD}
\end{eqnarray}
which consists of the two leads (1st term), single resonant QD
level (2nd term), and tunnelling between QD-$d$ and the leads
(last term). The operator $c_{\alpha k }$ ($c_{\alpha k}^{\dag}$)
annihilates (creates) an electron with energy $\varepsilon_{k}$ of
momentum $k$ on the lead $\alpha$. $d$ ($d^{\dag}$) annihilates
(creates) an electron in QD-$d$. QD-$d$ is modelled as a single
resonant level of its energy $\varepsilon_d$. A voltage $V_d$ is
applied across the detector which gives the difference in the
chemical potentials between the two leads by $eV_d$.
 The interaction between QD-$d$ and QD-$s$ is described by
$H_{int} = U d^{\dag}d\: (\hat{n}-1/2)$, where $\hat{n}$ and $U$
stand for the number operator for QD-$s$ and the inter-QD Coulomb
interaction, respectively. This interaction shifts the effective
energy level of the QD-$d$, which implies that the information of
the charge state of the QD-$s$ is {\em transferred} to the QD-$d$.
This transferred information enables detection of the charge state
in the QD-$s$. For convenience we introduce the $n$-dependent
energy level of the QD-$d$ where $n$ denotes the charge state of
the QD-$s$: $\varepsilon_n = \varepsilon_d  + (n-1/2) U$. Note
that the Hamiltonian of the `system' containing QD-$s$ is not
given explicitly since our main interest is to investigate the
detector. The effect of the scattering at the QDD is described by
the $n$-dependent scattering matrix $S_n$:
\begin{equation}
 S_n = \left( \begin{array}{cc}
                r_n & t_n' \\
                t_n & r_n'
           \end{array}  \right) \;. \label{Smtx}
\end{equation}
Here we have assumed that the QDD provides only a single
transverse channel.
 The unitarity of $S_n$ gives the constraints $|t_n|^2 +
|r_n|^2 = |t_n^{\prime}|^2 + |r_n^{\prime}|^2 =1$, and $r_n t_n^*
+ r_n^{\prime *}t_n^{\prime}=0 $.

In the absence of the external magnetic field, time reversal
symmetry (TRS) of the QDD is preserved, and thus
$t_n=t_n^{\prime}$. From the relation between the retarded Green's
function and the scattering matrix~\cite{langreth66,ng88}, one can
obtain the components of the scattering matrix given in the
following form describing the Breit-Wigner
resonance~\cite{breit36,landau77}:
\begin{eqnarray}
 r_n  =  \frac{\varepsilon - \varepsilon_n
    + i(\Gamma_L - \Gamma_R)}{\varepsilon - \varepsilon_n +
    i\Gamma},~
 r_n'  =  \frac{\varepsilon - \varepsilon_n
    - i(\Gamma_L - \Gamma_R)}{\varepsilon - \varepsilon_n +
    i\Gamma},~
 t_n &=& t_n' = \frac{-i2\sqrt{\Gamma_L \Gamma_R}}{\varepsilon -
\varepsilon_n + i \Gamma } \label{amplitudeRT},
\end{eqnarray}
where $\varepsilon$ is the incident energy of an electron and
$\Gamma = \Gamma_L + \Gamma_R $ with $\Gamma_\alpha$
($\alpha=L,R$) being the coupling strength between the lead
$\alpha$ and QD-$d$ given by $\Gamma_\alpha = \pi N_{\alpha} (0)
|V_{\alpha}|^2$, where $N_{\alpha}(0)$ denotes the density of
states at the Fermi level of the lead $\alpha$.

The information of the charge state in QD-$s$ is transferred into
 the detector through the transmission probability
 $|t_n|^2$ and the relative scattering phase $\phi_n = \arg(t_n/r_n)$.
The `measurement' is performed by the transmission probability
change of the detector due to an extra electron in the QD-$s$. The
measurement rate is defined by \cite{aleiner97,korotkov01}
\begin{equation}
 \Gamma_{m} = \frac{eV_d}{h}  \frac{(\Delta T)^2}{4T(1-T)}~,~~~~~~~ \label{GM}
\end{equation}
where $ T =( |t_1|^2 + |t_0|^2 )/2$ and $\Delta T =
|t_1|^2-|t_0|^2 $. Concerning the phase information $\phi_n$, it
is not actually measured. Therefore the phase does not affect the
measurement rate. However it is related to dephasing because there
is a {\em possibility} to measure the phase regardless of whether
it is being measured or not~\cite{stern90,scully91}.  In a
symmetric QDD  ($\Gamma_L=\Gamma_R$), one can find that from Eq.
(\ref{amplitudeRT}) $ \phi_n = - \pi/2 $ for $ \varepsilon <
\varepsilon_n$
 and $ \phi_n = + \pi/2 $ for $ \varepsilon > \varepsilon_n$.
In contrast to the non-resonant detector, there {\em is} a phase
jump by $\pi$ at $\varepsilon=\varepsilon_n$. This phase jump
contributes to the dephasing rate, and therefore distinguishes the
dephasing from the measurement rate.

In order to describe the hybrid system, we adopt the density
matrix formulation~\cite{Gurvitz97,hacken98,hacken01}. combined
with the scattering matrix. The effect of the capacitive
interaction between the QDD and the QD-$s$ is described by a
two-particle scattering matrix \cite{hacken98,hacken01}
$\mathbf{S}$  where its elements are given by
\begin{equation}
 S_{nn^\prime}= \delta_{nn^{\prime}} ( \delta_{n0}S_0 +
\delta_{n1} S_1 ), ~~(n,n^{\prime}\in \{0,1\}).
\end{equation}
In the case of a single scattering event in the QDD with  the
initial state of the total system $ |\psi^{0}_{tot}\rangle =
(a|0\rangle + b|1\rangle )\otimes |\chi_{in}\rangle$ where  the
initial state of QD-$s$ is coherent superposition of the $n=0$ and
$n=1$ states denoted by $a|0\rangle + b|1\rangle $ with
$|a|^2+|b|^2 = 1$ and the initial state of the QDD is $
|\chi_{in}\rangle$ with incident electron from the lead $L$, the
state of QD-$s$ after the scattering is described by the reduced
density matrix $\rho = {\rm Tr}_{\mathsf{QDD}} \{\mathbf{S}
   |\psi_{tot}^0\rangle \langle\psi_{tot}^0| \mathbf{S}^{\dag}  \}$.
The reduced density matrix $\rho$ is given by
 $ \rho = |a|^2 |0\rangle \langle 0| +\lambda ab^{*} |0\rangle \langle 1|
  +\lambda^{*} a^{*}b |1\rangle \langle 0| + |b|^2 |1\rangle \langle
  1|$,
 where $ \lambda = r_0 r_1^{*} + t_0 t_1^{*}$. One can find that
the diagonal elements of $\rho$ do not change upon scattering, but
the off-diagonal elements are modified by $ \rho_{01} = \lambda
\rho_{01}^0~~{\rm and}~~ \rho_{10} = \lambda^* \rho_{10}^0  $.
%,\label{densitymatrixofAB}
We consider the limit where the scattering in the QDD occurs on a
time scale much shorter than the relevant time scales in the
QD-$s$. In the present case, $\Delta t \ll t_d$, where $\Delta t=
h/2eV_d$ is the average time interval between two successive
scatterings, and $t_d$ is the dephasing time of the charge state
of the QD-$s$ induced by the QDD. In this limit, one can find that
the time evolution of $\rho_{01}$ is given as~\cite{hacken01}
\begin{equation}
\rho_{01}(t) = e^{-(\Gamma_d -i\nu)t}\rho_{01}^{0},
\end{equation}
where $\nu  = \arg(\lambda)/\Delta t $, and
\begin{equation}
 \Gamma_d = \frac{1}{t_d} =
  -\frac{\ln |\lambda|}{\Delta t}= -\frac{2eV_d}{h}\ln |\lambda|
 ~. \label{Gammad-lambda}
\end{equation}

In the weak measurement limit  ($\lambda\ \approx 1 $), the
dephasing rate $\Gamma_d$ can be expanded in terms of the change
of the transparency $\Delta T = |t_1|^2 - |t_0|^2$ and the change
of the relative scattering phase $\Delta \phi = \arg(t_1/r_1) -
\arg(t_0/r_0) $ as follows
\begin{eqnarray}
\Gamma_d = \Gamma_T + \Gamma_\phi ~{\rm with}~
 \Gamma_{T} = \frac{eV_d}{h}  \frac{(\Delta T)^2}{4T(1-T)}
 ~{\rm and}~
  \Gamma_\phi = \frac{eV_d}{h}T(1-T)(\Delta \phi)^2 .
  \label{ExpansionGdGTGphi}
\end{eqnarray}
One can find that $\Gamma_T=\Gamma_m$: That is, the current
sensitive component of the dephasing rate is equivalent to the
measurement rate in Eq. (\ref{GM})  ~\cite{note1}.

The detector efficiency, namely $\eta$, is defined by the ratio
between the measurement rate ($\Gamma_m$) of the detector and the
dephasing rate as~\cite{korotkov01,pilgram02}
\begin{equation}
 \eta \equiv \Gamma_m/\Gamma_d \;.
\end{equation}
For a QPC detector that obeys the TRS and the mirror reflection
symmetry (MRS), it has been shown that the phase-sensitive
dephasing does not take place because the relative phase between
the transmission and the reflection amplitudes remains constant
(that is $\Delta\phi=0$)~\cite{korotkov01,pilgram02}. Therefore
the efficiency for a symmetric single-channel detector has its
maximal value independent of the transparency of the detector.
 In the following, based on a symmetry argument for the scattering
matrix, we show that the phases of the scattering amplitudes may
play an important role in the detector efficiency in spite of the
TRS and the MRS.

%%%%%%%%%%%%%%%%%%%%%%%%%%%%%%%%%%%%%%%%%%%%%%%%%%%%%%%%%%%%%%%%%%%%%%%%%%%5
\begin{figure}[b]
  \resizebox{7.5 cm}{!}{\includegraphics[26,40][351,205] {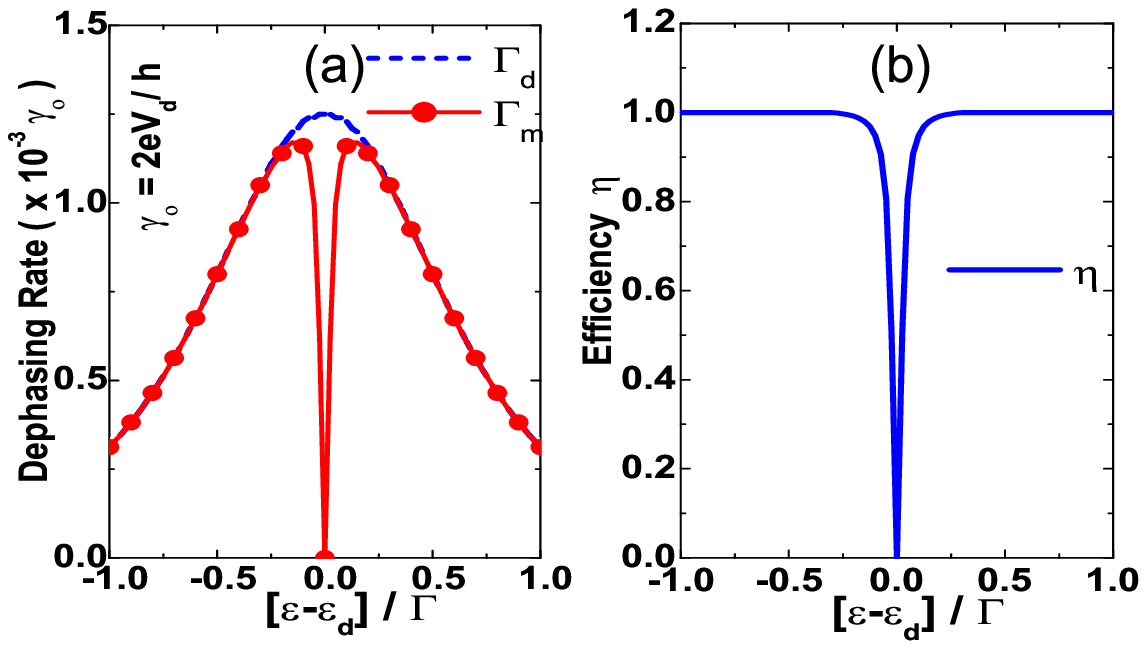}    }
  $~~~$\resizebox{7.5 cm}{!}{\includegraphics[26,40][351,205] {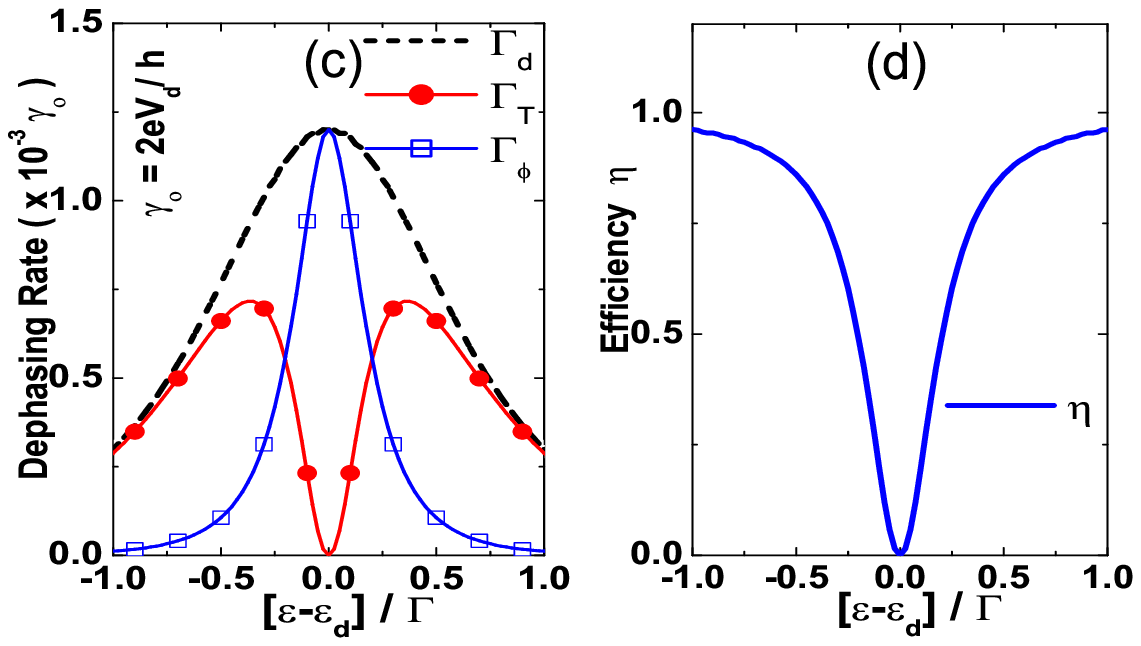} }

\caption{\label{symmetric} (a) The measurement rate ($\Gamma_m$)
and the dephasing rate ($\Gamma_d$), and (b) the detector
efficiency, for a symmetric QDD ($\Gamma_L = \Gamma_R = 0.5
\Gamma$).
 (c) The  dephasing rates $\Gamma_d$,
$\Gamma_T$, $\Gamma_\phi$, and (d) the detector efficiency, for an
asymmetric QDD ($\Gamma_L=0.4\Gamma, \Gamma_R=0.6 \Gamma$).
 $U =0.05 \Gamma$. $ \gamma_0=2eV_d/h$.}
\end{figure}
%%%%%%%%%%%%%%%%%%%%%%%%%%%%%%%%%%%%%%%%%%%%%%%%%%%%%%%%%%%%%%%%%%%%%%%%%%%

Concerning the relation between the dephasing and the symmetry of
the detector, we reexamine the scattering matrix for the detector.
In general, a scattering matrix for a single-channel transport can
be written in the following form~\cite{hwlee99,taniguchi99}:
\begin{equation}
 S =  \left(
        \begin{array}{cc}
           r & t' \\
       t & r'
     \end{array}
       \right)
   = e^{i\theta} \left(
                 \begin{array}{cc}
          \sqrt{R} e^{i\varphi_1} & i\sqrt{T} e^{-i\varphi_2} \\
          i\sqrt{T} e^{i\varphi_2} & \sqrt{R} e^{-i\varphi_1}
         \end{array}
                    \right) \;,
\end{equation}
with the constraint $R+T=1$. In the presence of the TRS ($t=t'$)
and the MRS ($r=r'$), the phase components $\varphi_1$ and
$\varphi_2$ should satisfy $\varphi_1=n_1\pi$ and
$\varphi_2=n_2\pi$, respectively, where $n_1$ and $n_2$ are
integers. Therefore, the components of the scattering matrix can
be written as follows:
\begin{eqnarray}
 r &=& r' = \sqrt{R} e^{i\theta}\;\;\; \mbox{\rm or}\;\;\;
    -\sqrt{R} e^{i\theta} \:, \\
 t &=& t' = i\sqrt{T} e^{i\theta} \;\;\; \mbox{\rm or}\;\;\;
    -i\sqrt{T} e^{i\theta} \;.
\end{eqnarray}
 From these, one can find that $\arg(t/r)=\pm\pi/2$, implying $\Delta\phi=0$ or $\Delta\phi = \pi$.
 {\em The latter has not been noticed previously}.
 Indeed we have shown in the previous section that the scattering
matrix for a QDD (Eq.~(\ref{Smtx},\ref{amplitudeRT})) satisfies
this condition for $\Gamma_L=\Gamma_R$ (symmetric QDD). For
$\Gamma_L= \Gamma_R$, $\Delta\phi=0$ except at resonance
($\varepsilon=\varepsilon_d$) where the abrupt change of the
relative scattering phase ($\Delta\phi=\pi$) takes place due to
the fact that $r=0$. For this reason, the formula for a weak
measurement limit of Eq. (\ref{ExpansionGdGTGphi}) is not valid
for a symmetric detector since the phase-sensitive term cannot be
well defined.

Fig.~2 shows (a) the dephasing rate $\Gamma_d$ (calculated from
Eq. (\ref{Gammad-lambda})) and the measurement rate $\Gamma_m$
(calculated from Eq.~(\ref{GM})), and (b) the efficiency $\eta$
for a symmetric QDD. In contrast to the single-channel
non-resonant detector, the efficiency is not always at its maximum
value of 1 even in the presence of the TRS and the MRS. Instead,
it displays a dip around the resonance. This behavior originates
from the abrupt phase change of scattering phase at resonance.
According to Eq.~(\ref{amplitudeRT}), the phase of the reflection
amplitude for a symmetric QDD ($\Gamma_L=\Gamma_R$) changes
abruptly from $+\pi/2$ to $-\pi/2$. This abrupt phase change
causes another source of dephasing insensitive to the detector
current, and accordingly, lowers the efficiency of the detector.
For an asymmetric detector ($\Gamma_L\ne\Gamma_R$),
Eq.~(\ref{ExpansionGdGTGphi}) can be used in the weak measurement
limit since scattering phases do not have discontinuity. Fig.~2
shows (c) the dephasing rates $\Gamma_d,\Gamma_T,\Gamma_\phi$, and
(d) the efficiency of the detector. Near the resonance the
current-sensitive dephasing ($\Gamma_T$) shows a dip, which
reduces the detector efficiency. On the other hand, the
phase-sensitive dephasing has a peak around the resonance. As the
asymmetry increases, the dip width of $\Gamma_T$ and the peak
width of $\Gamma_\phi$ increase. In fact, these widths are
proportional to the degree of the asymmetry,
$|\Gamma_L-\Gamma_R|$.

In the weak measurement limit ($|\lambda|\sim1$), from the
relation $\ln|\lambda| \cong 1 - |\lambda|$,
Eq.~(\ref{amplitudeRT}), and Eq. (\ref{Gammad-lambda}), we obtain
\begin{equation}
 \Gamma_d =  \frac{eV_d}{h} \frac{ 4\Gamma_L\Gamma_R  U^2 }
     {[(\varepsilon-\varepsilon_d)^2 + \Gamma^2 ]^2}
  = \frac{eV_d U^2 }{4h\Gamma_L\Gamma_R} T^2~. \label{simT2}
\end{equation}
This result is in very contrast with the non-resonant detector in
the following aspects. First, the dephasing rate for a QDD
increases as $T$ increases, while for a QPC detector, it has a
maximum value in the intermediate value of $T\simeq 1/2$ and
vanishes at the two extrema $T=0$ and $T=1$. One may understand
that $\Gamma_d$ has its maximum at resonance, since a QDD is a
detector based on the charge sensitivity of the resonant
tunnelling. As we discussed above, although the current is
insensitive to the charge state of the QD-$s$ at resonance,
discontinuity (for a symmetric QDD) or rapid change (for an
asymmetric QDD) of the scattering phase {\em stores} the
information of the quantum state of the QD-$s$, thereby induces
strong dephasing. Second, Eq.~(\ref{simT2}) clearly shows that the
dephasing rate has nothing to do with the shot noise of the
detector. As pointed out before~\cite{levinson97}, the dephasing
rate is formally different from the shot noise of the detector.
The difference originates from the fact that the dephasing is
caused by the charge fluctuations but the shot noise comes from
the current fluctuations of the detector. However, QPC detectors
mostly show that the dephasing rate is proportional to the shot
noise of the detector (See Ref.~\cite{kang05} and references
therein). An interesting point of our study is that the QDD
dramatically shows that the two quantities are completely
independent.

Next we consider the case where a QDD contains background
transmission as well as resonant tunnelling. In this case
transport through the QDD shows the Fano resonance. The QDD is
assumed to have TRS and MRS. The Hamiltonian for the QDD is given
by $ \tilde{H}_d = H_d + \sum_{k} \big( W_{LR} c^{\dag}_{R k} c_{L
k } + h.c. \big)$, where the first term describes the resonant
tunnelling (given by Eq.~(\ref{QDD})) and the second term
represents the direct (non-resonant) transmission between the two
leads. Following the procedure of calculating the retarded Green's
function and the components of the scattering
matrix~\cite{kang05a,ng88,langreth66}, the transmission ($\tilde
t_n, \tilde t_n'$) and the reflection ($\tilde r_n, \tilde r_n'$)
coefficients can be calculated. For the symmetric case $\Gamma_L
=\Gamma_R=\Gamma$,
 we get \label{eq:s-Fano}
\begin{eqnarray}
 \tilde r_n ~=~ \tilde r_n^{\prime} = \frac{\sqrt{1-T_b}(\varepsilon -
   \tilde\varepsilon_n) + \sqrt{T_b  }\tilde\Gamma }
    { \varepsilon -\tilde\varepsilon_n - i {\tilde\Gamma} }~,~~
 \tilde t_n ~=~ \tilde t_n^{\prime} = i\frac{\sqrt{T_b}(\varepsilon -
   \tilde\varepsilon_n ) - \sqrt{1-T_b}\tilde\Gamma }
    { \varepsilon-\tilde\varepsilon_n - i {\tilde\Gamma} }~.\label{Fanort}
\end{eqnarray}
where $T_b$ corresponds to the probability of the background
transmission. The dot energy level and the resonance width are
renormalized due to the background transmission as
$\tilde\varepsilon_n=\varepsilon_n-\kappa\tilde\Gamma$ (or
$\tilde\varepsilon_d=\varepsilon_d-\kappa\tilde\Gamma$), and
$\tilde\Gamma = \Gamma/(1+\kappa^2)$, respectively. Here the
parameter $\kappa$ is defined as $\kappa=\pi N(0)|W_{LR}|$ where
$N(0)$ denotes the density of states of a lead. In terms of the
parameter $\kappa$, the background tunnelling probability is
written as $T_b = 4 \kappa^2/(1+\kappa^2)^2 $.

\begin{figure}[b]
\begin{center}
 \resizebox{7.5 cm}{!}{\includegraphics[26,40][351,205] {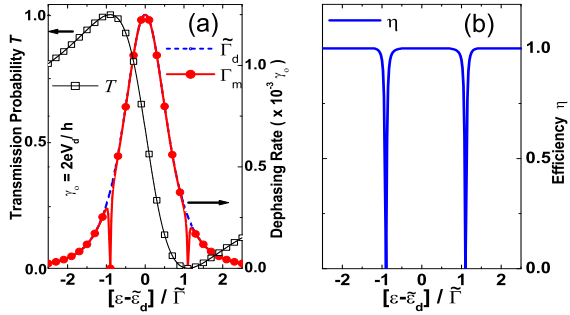} }
\caption {\label{Fano} (a) The transmission probability $T
=(|\tilde t_0| +|\tilde t_1| )/2$, the measurement rate
$\Gamma_m$, and the dephasing rate $\tilde\Gamma_d$, and (b) the
detector efficiency $\eta $, for a symmetric QDD with background
transmission probability $T_b=0.45$, and $U = 0.05 \tilde\Gamma$.}
\end{center}
\end{figure}

 In the limit of $T_b = 0$ (that is, in the absence of the background
transmission), Eq.(\ref{Fanort}) is equivalent to the result
obtained in Eq.(\ref{amplitudeRT}), as one can expect.
 Eq.~(\ref{Fanort}) shows that the relative scattering phase
($\arg(\tilde t_n/ \tilde
 r_n)$) changes abruptly by $\pi$ both at the reflection zero and at the
transmission zero, where the measurement rate $\Gamma_m$ vanishes.
On the other hand, the dephasing rate $ \tilde\Gamma_d$ via a QDD
with Fano resonance is written as
\begin{equation}
 \tilde\Gamma_d =  \frac{eV_d}{h} \frac{\tilde\Gamma^2  U^2 }
 {[(\varepsilon -\tilde\varepsilon_d)^2 +\tilde \Gamma^2 ]^2} ~.
  \label{FanogammaDeltaU}
\end{equation}
This shows that the dephasing rate is not modified by the presence
of the background transmission (see Eq.~(\ref{simT2})), except the
renormalization of the energy level and the resonance width. Thus,
one can conclude that the characteristics of the dephasing is not
affected by the non-resonant transmission component, aside from
the renormalization of the parameters, while the conductance
(which is proportional to $|\tilde{t_n}|^2$) is severely modified.
Fig.~3 displays (a) the dephasing rate $\tilde\Gamma_d$ and the
measurement rate $\Gamma_m$ together with the transmission
probability $T$, and (b) the detector efficiency, for a symmetric
QDD with the background tunnelling probability $T_b=0.45$. As one
can see, $\tilde\Gamma_d$ is equivalent to what is expected
without the background transmission. One peculiar feature is the
two dips in $\Gamma_m$ (and also in the detector efficiency
$\eta$): one from the transmission zero, the other from the
reflection zero.

In conclusion, the dephasing rate of a mesoscopic system via a QDD
is maximum at resonance of the QDD where the conductance is
insensitive to the charge state of the mesoscopic system.  As a
result, the efficiency of the detector shows a dip structure and
vanishes at resonance, in contrast to the symmetric non-resonant
detector which retains a maximum detector efficiency. The anomaly
of the QDD originates from the abrupt phase change of scattering
amplitudes in the presence of resonance, which is insensitive to
the detector current but {\em stores} the information of the
quantum state of the mesoscopic system.  If the QDD shows Fano
resonance, there are two dips in the detector efficiency which
correspond to the transmission and the reflection zeros.

\acknowledgements%
This work was supported by the Korea Research Foundation
(R05-2004-000-10826,  KRF-2005-070-C00055), the Korean Ministry of
Science and Technology, and Chonnam National University Grant
(2005).


\begin{thebibliography}{0}
\bibitem{note1} The two quantities $\Gamma_m$ and $\Gamma_T$  might be
 different by a constant factor of {\cal O}(1), depending on the
 definition of $\Gamma_m$. See Ref.~\cite{korotkov01}.
\end{thebibliography}
\end{document}